\documentclass[12pt,twoside,a4paper,useAMS]{meteoroids2013}
\usepackage{color,graphicx,float}
\usepackage[english]{babel}
\usepackage[round]{natbib}
\title[ADAM-WFS] %% give here short title %%
{Automatic Detection of Asteroids and Meteoroids - A Wide Field Survey} %%full title

\author[Vere\v{s} et al.]   %% give here short author list %%
{Vere\v{s} P.$^{1,2}$,
 T\'{o}th J.$^2$, Jedicke R.$^1$, Tonry J.$^1$, Denneau L.$^1$, Wainscoat R.$^1$, Korno\v{s} L.$^2$ %\break
 \and \v{S}ilha J.$^{2,3}$}

\affiliation{$^1$Institute for Astronomy, University of Hawaii at Manoa, 2680 Woodlawn Drive, Honolulu, HI 96814, USA (email: veres@ifa.hawaii.edu)
\\[\affilskip]
$^2$Faculty of Mathematics, Physics and Informatics, Comenius
University, Mlynska Dolina, 84248 Bratislava, Slovakia
\\[\affilskip]
$^3$Astronomical Institute, University of Bern, Sidlerstrasse, CH-3012 Bern, Switzerland}

\pagerange{110--116}
\jname{Proceedings of the Meteoroids 2013 Conference\\
       Aug. 26-30, 2013, A.M. University, Pozna\'{n}, Poland}
\editors{Jopek T.J., Rietmeijer F.J.M., Watanabe J., Williams I.P., ed.}

\begin{document}
\maketitle
\begin{abstract}
We propose a low-cost robotic optical survey aimed at $1-300$ m Near Earth Objects (NEO) based on four state-of-the-art
telescopes having extremely wide field of view. The small
Near-Earth Asteroids (NEA) represent a potential risk but also
easily accessible space resources for future robotic or human space in-situ
exploration, or commercial activities. The survey system will be
optimized for the detection of fast moving - trailed - asteroids,
space debris and will provide real-time alert notifications. The
expected cost of the system including 1-year development and
2-year operation is 1,000,000 EUR. The successful demonstration of
the system will promote cost-effectiveicient ADAM-WFS (Automatic
Detection of Asteroids and Meteoroids - A Wide Field Survey)
systems to be built around the world.

\end{abstract}
\section{Introduction}
It has been more than 200 years since the first asteroid, now
defined as a dwarf planet, Ceres, was discovered. Progressive
development of instrumentation and techniques in astronomy revealed the existence
of the asteroid main belt and other dynamically stable and unstable
populations of minor bodies throughout the Solar System including
NEAs encountering the Earth. In 2013, 10,000th NEO was discovered
by the Pan-STARRS survey \citep{Kaiser10}. Although, the number of NEA has been
rising rapidly within the last decade due to the dedicated
asteroid surveys such as Spacewatch \citep{Gehrels98}, LINEAR \citep{Stokes00}, LONEOS \citep{Koehn99}, NEAT \citep{Pravdo99},
space-based NEOWISE \citep{Wright10} or ongoing next generation surveys such as
Catalina Sky Survey \citep{Larson98} and Pan-STARRS, there is a large uncertainty in the population count
and orbital properties of small NEA within the size range of
$1-300$\,m are not understood well. In previous studies
(\cite{Rab14}, \cite{Bot02} and \cite{Stu04}) the population count
of 10\,m size NEA differed more than one order of magnitude.
Although, NEOWISE mission supposedly derived accurate diameters of
asteroids by the thermal modeling and assumed that there are less
small NEAs that it was predicted before (\cite{Mai13}), recent
studies by \cite{Har08} and \cite{Bro13}, based on recent
ground-based discoveries and the fall of the Chelyabinsk meteoroid
in 2013, suggest a much higher count. Other previous studies
also indicated that small NEA population could be enhanced by tidal disruption
of rubble-pile asteroids during close approaches to the Earth, such as a theory about the
common origin of P{\v r}{\'i}bram and Neuschwanstein meteorites coming from heterogenous stream of meteoroids (\cite{Spu03}) when
the frequency of rubble-pile asteroids disruption were
estimated \citep{Tot11}. Similarly in subsequent studies, \cite{Sch12}, \cite{Sch14} analysed
the creation mechanisms of NEA's families and showed supporting evidence for mentioned theories. Moreover, the meteoroid streams may contain large
particles from break-ups and enriched population of small NEOs
(\cite{Por92}, \cite{Rud12}, \cite{Bab13}).

In the past (\cite{Tot02}, \cite{Tot03}, \cite{Ver06}), we
proposed a simple low-cost survey for discovering population of
small asteroids flying-by within one lunar distance from the
Earth. In this paper we introduce an advanced and more
sophisticated, yet low cost concept that will characterize the
population of small NEA that from the large part is undetectable
by current telescopic systems.

\section{Concept}
\begin{figure}[!ht]
 \centering
 \includegraphics[width=130mm]{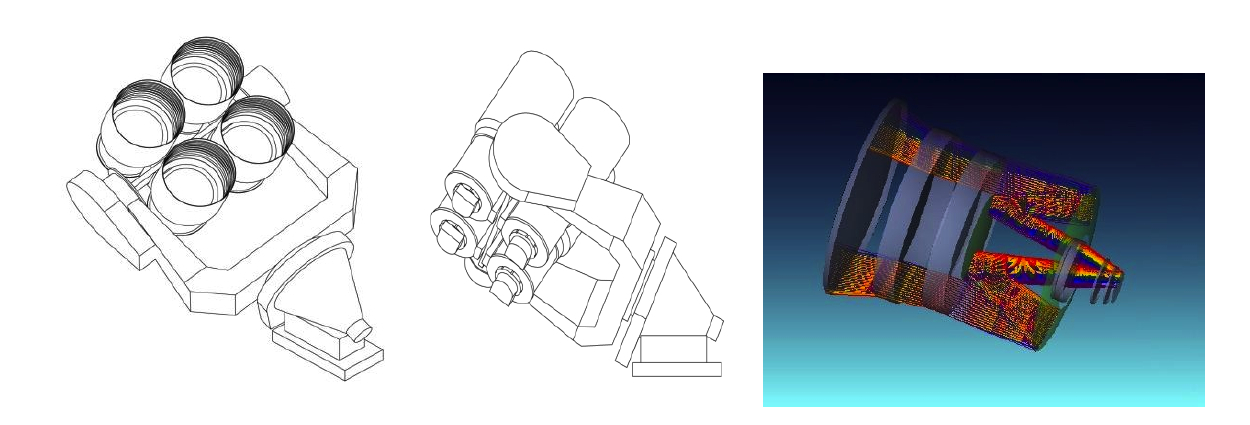}
 \caption{A design of the ADAM survey telescope - the mount and 4 optical assemblies in front and rear views (left) and optical path of the Houghton-Terebizh optical system (right).}
 \label{fig:design}
\end{figure}

\begin{figure}[!ht]
 \centerline{\includegraphics[width=10.0cm]{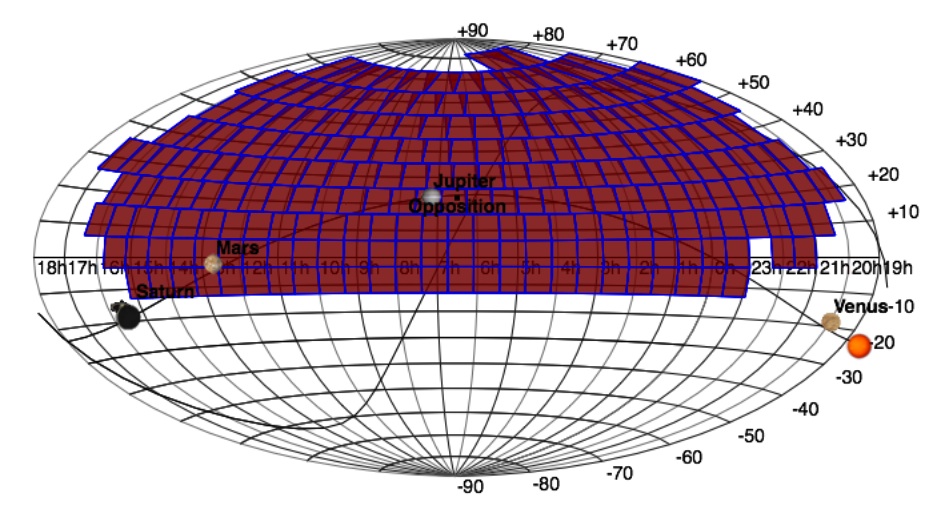}}
 \caption{Sky coverage of the ADAM survey on one night, taken from the MOPS simulation interface.}
 \label{fig:sky}
\end{figure}

Automatic Detection of Asteroids and Meteoroids - A Wide Field
Survey (ADAM-WFS) will consist of 4 identical wide-field \citep{Ter11}
astrographs (Houghton-Terebizh D=300 mm, f/1.44) on a fast-track
mount with high-precision guiding (Fig. ~\ref{fig:design}). Each
telescope will be equipped with a large-scale single chip CCD
camera (4096x4096 pix) providing a total FOV of almost 100 square
degrees. The predicted limiting magnitude with the wide-band
optical filter will be +17.5 mag. at S/N=5.0 with 30 sec exposures
and a pixel scale of 4.36 arcsec/pix. This configuration is able
to survey almost an entire sky visible from a specified location
in 3 visits per night (Fig. ~\ref{fig:sky}), with the rapid image
processing providing moving targets in almost a real-time. We will
use the Moving Object Processing System (MOPS, \cite{Den07}) that
has been utilized by the Pan-STARRS and ATLAS \citep{Tonry11}. Stationary transients
will be processed during the daytime. We propose to build the
system at an existing observatory with a dedicated 60-80 cm
follow-up telescope and existing infrastructure.

\subsection{Advantages of ADAM-WFS}

The survey budget is considered \textbf{low-cost} compared to the
existing or planned all-sky surveys that focus on deeper limiting
magnitude, or space-based observatories, which are up to two
orders of magnitude more expensive. In contrast to existing or
planned deep surveys with narrower fields of view, our survey will
cover \textbf{entire visible night sky} few times per night. In
spite of its lower limiting magnitude, large pixels would
\textbf{decrease} the effects of \textbf{trailing loss} for fast
moving targets. Thus, ADAM-WFS will detect more fast moving objects
than any survey equipped with telescopes of similar size.

Usually, the development of a new survey takes years due to the new
hardware and software development and methods to be implemented.  Our
goal is to \textbf{avoid reinventing the wheel} and use existing
routines for image processing, moving object processing, hardware,
mount and optics to speed-up the delivery of the complete
system and cut down the cost. We will also count on a
\textbf{compact team} of astronomers and engineers with work and
science experience on existing surveys (Pan-STARRS) or surveys
under the development (ATLAS), on full-time and
sub-contracts.

Significant advantage is the \textbf{existing infrastructure} that
will serve the prototype for the development and operation. It
will be built on existing observatory -- Astronomical and
Geophysical Observatory of the Comenius University, Modra,
Slovakia (AGO Modra) -- that contains workshop, complete infrastructure and
nonstop technical support. Future deployment of ADAM-WFS systems
is also strongly encouraged on existing observatories (e.g. Canary
Islands, South Africa). New generation surveys are often focused
on multiple tasks (e.g. Pan-STARRS). Our survey will be
\textbf{optimized towards small NEA detection} with byproducts
available for additional science.

\subsection{Targets of the survey}
\begin{figure}[!b]
 \centering
 \includegraphics[width=6.5cm]{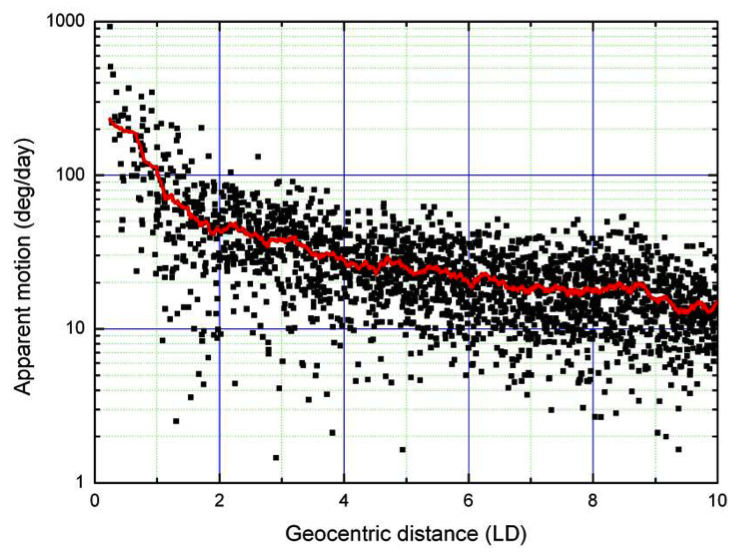}
  \includegraphics[width=6.5cm, height=5.1cm]{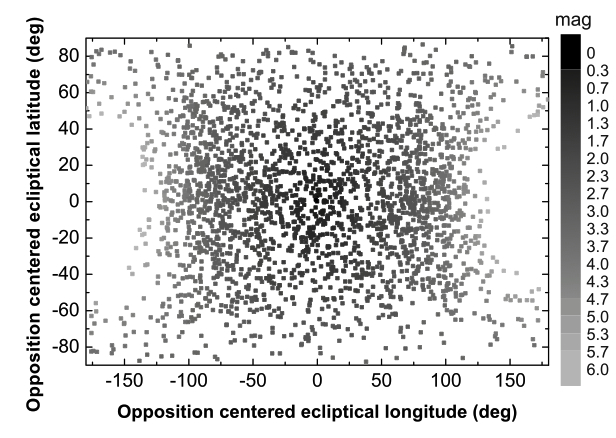}

  \caption{ Rate of motion of asteroids on near-miss orbits as a
function of the minimum geocentric distance denoted in lunar distances (left) and magnitude loss (a
phase effect) of asteroids as a function of opposition centric
ecliptical distance.}
 \label{fig:asteroids}
\end{figure}

The main target of the survey is \textbf{discovery and
characterization of small NEA and other close-approaching
populations}. The survey will search for potential \textbf{Earth
impactors} of small and intermediate diameters (1--100\,m) and
\textbf{pre-entry detections of bright bolides}. The all-sky
coverage and optimization toward high angular rate of motion would
make the system a good detector for \textbf{monitoring,
characterization and discovery of space debris}. This combination
of survey properties will also benefit the detection of
\textbf{telescopic meteors} as well. Because the limiting
magnitude will not be a competition for multi-meter apertures and
space based telescopes, it will serve the photometry of objects
that are too bright and saturated for next generation surveys,
such as \textbf{ bright main belt asteroids}. The extremely low
focal ratio and large light gain of the aperture would help
\textbf{discovery of active asteroids} (\cite{Jew12}), especially
at low solar elongations.  The important feature of the project
will be the accessibility of the data - the database will be
available for external scientists to mine additional resources for
\textbf{science byproducts}, such as \textbf{variable stars,
novae, supernova, lensing events, gamma-ray bursts}.

\begin{figure}[t]
 \centering
 \includegraphics[width=0.8\textwidth]{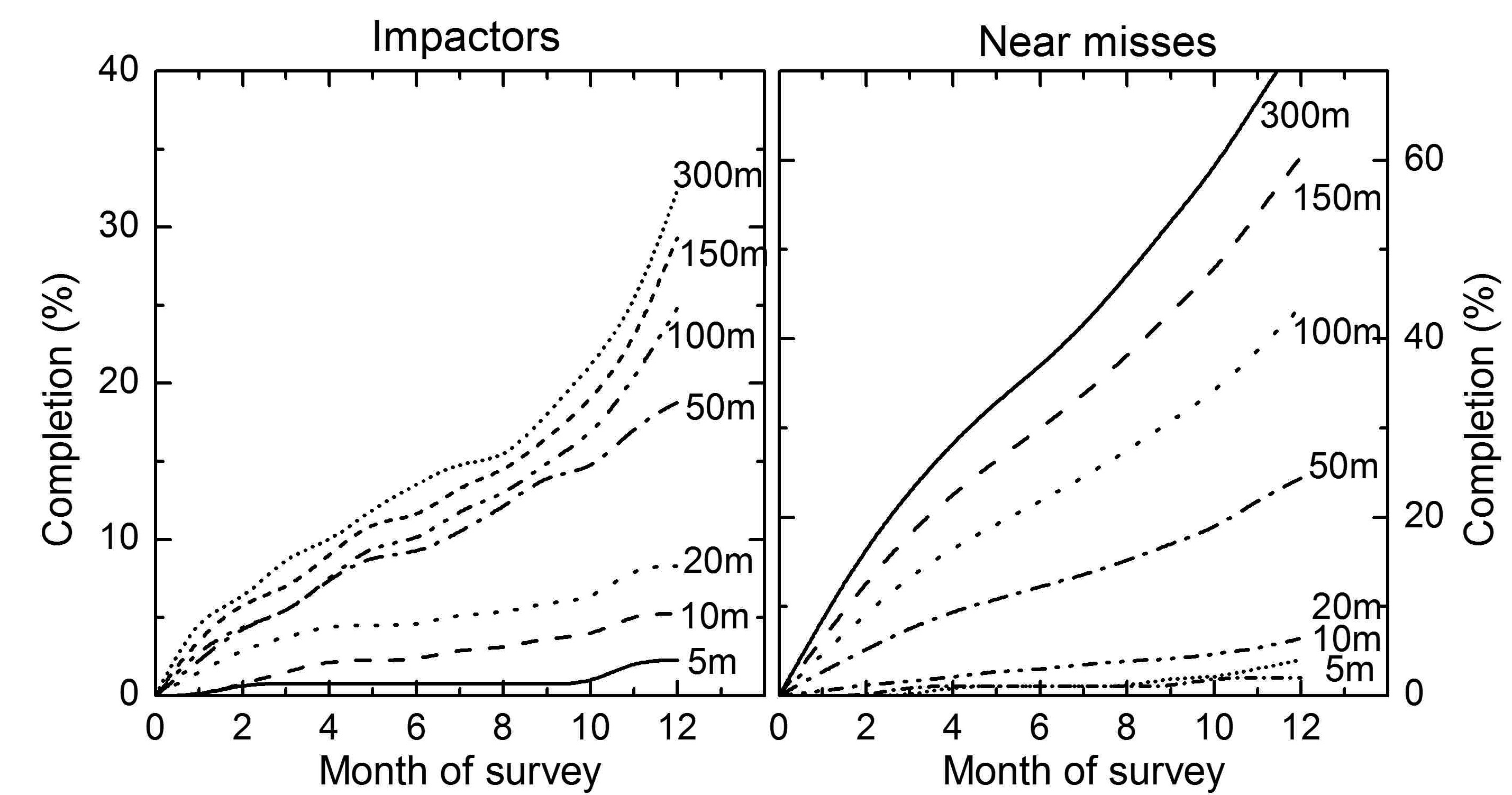}
 \caption{Survey efficiency in finding Earth impacting asteroids
(left) and asteroids that fly-by in the Earth vicinity (right).}
 \label{fig:eff}
\end{figure}

\section{Expected outcomes}
\begin{figure}[t]
 \centering
 \includegraphics[width=0.7\textwidth]{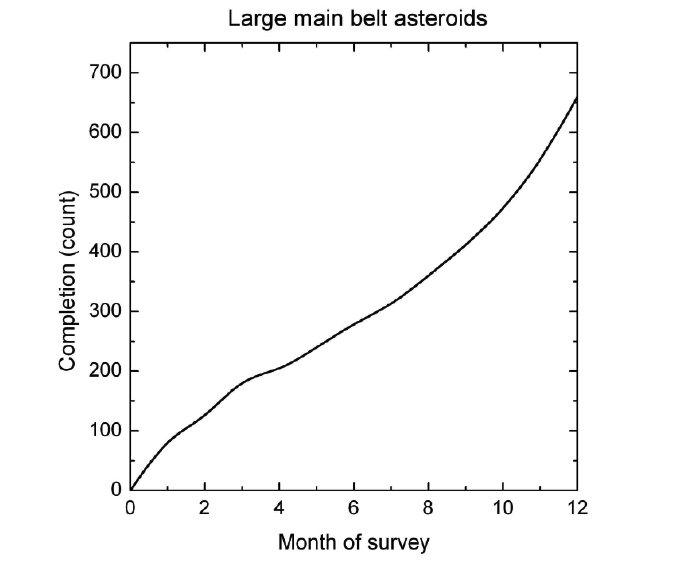}
 \caption{Survey efficiency of large main belt asteroids.}
 \label{fig:effMB}
\end{figure}

\begin{figure}[!hb]
 \centering
 \includegraphics[width=0.8\textwidth]{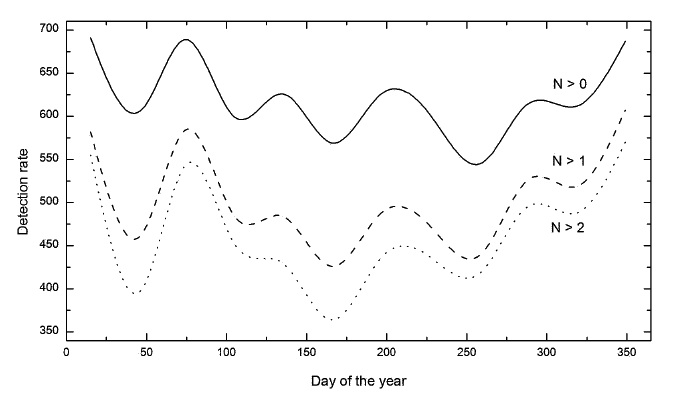}
 \caption{Number of detection of space debris per night by ADAM-WFS
based on space-track data. The number of detections per particle
per night ($N>0$, at least one detection; $N>1$, two and more
detections; $N>2$, more than two detections). Variations in the
rates are produced by the combination of weather conditions and
Earth's shadow vs. field of view geometry.}
 \label{fig:debris}
\end{figure}

We simulated a one-year ADAM-WFS survey based at AGO Modra
Observatory, by using MOPS with the realistic pointings, avoiding
Moon and using orbits of large MB asteroids, synthetic orbits of
asteroids that will approach the Earth within 10 lunar distances
based on real asteroids and Earth impacting asteroids
\citep{Ches04,Den13}. The apparent rate of motion of
asteroids at the closest distance and phase effect of asteroids
near the Earth are shown in Fig. \ref{fig:asteroids}. Figure
\ref{fig:eff} shows the efficiency of the proposed system in
discovering Earth impacting asteroids and close approachers as the
function of diameter and the duration of the survey. Depending on
the population model, this system will be able to discover
$30-120$ NEAs with $D>10\,m$ within 10 lunar distances per year
and a comparable number of smaller asteroids with diameter
$D<10\,m$. Figure \ref{fig:effMB} shows the number of known large
main belt asteroids that will be discovered within one year
survey. Due to the cadence of the survey we will obtain $\sim650$
light  curves of bright main belt asteroids every year, suitable
especially for slow rotators detection. We also performed the
simulation of space debris detection by using the SPACE-TRACK
debris catalog. The simulation detected 350 - 550 space debris
particles per night (Fig.~\ref{fig:debris}).

\section{Conclusion}

The project of ADAM-WFS represents a new way of observation and exploration of
the small NEO population in the close vicinity of the Earth. The
project will fill up the current gap in our knowledge of small
solar system bodies between the bolide-sized objects observed in the
atmosphere and the large asteroids and comets observed telescopically. Identical telescopes can be installed all over the world or cooperate
with other similar projects like ATLAS to increase the sky coverage and not depend on the daylight cycle to search for small NEOs, Earth
impactors and optical transient events like variable stars, novae
or supernovae. Data gathered during the operation will provide
terabytes of images and database entries for years of research and
data mining. Naturally, additional coordinated follow-up
observations will be needed to complete the orbital and physical determination of newly found asteroids.

\begin{acknowledgments}
We thank for the financial support from NASA grant No. NNX12AR65G
and Slovak Research and Development Agency Grant No. APVV 0516-10
and APVV 0517-12.
\end{acknowledgments}

\end{document}